\documentclass{mpe_report}

\usepackage{psfig,graphicx,epsfig}
\usepackage{color}
\usepackage{amsmath,amssymb,epic,eepic,array}
\usepackage{natbib}

\begin{document}

\pagenumbering{arabic}
\setcounter{page}{88}

 \renewcommand{\FirstPageOfPaper }{ 88}\renewcommand{\LastPageOfPaper }{ 91}

\title{What is special about High Magnetic Field Radio Pulsars?}
\subtitle{-- First results from the multi\-frequency polarimetry}
\author{Nata\v{s}a Vrane\v{s}evi\'c\inst{1,2}, Richard N. Manchester\inst{1}, 
Donald B. Melrose\inst{2}}
\institute{Australia Telescope National Facility, CSIRO, PO Box 76, Epping NSW 
1710, Australia 
\and  School of Physics, University of Sydney, Sydney NSW 2006, Australia}
\authorrunning{N. Vrane\v{s}evi\'c, R. N. Manchester \& D. B. Melrose}
\titlerunning{What is special about HBRPs}
\maketitle

\begin{abstract}
The Parkes Multibeam Survey led to the identification of a number of
long-period radio pulsars with magnetic field well above the `quantum
critical field' of $\sim$ 4.4 $\times $10$^{13}$ G
(HBRPs). The HBRPs have similar spin parameters to magnetars,
 but their emission properties are different, and contradict
those theories that predict that radio
emission should be suppressed above this critical field.  
Our observations support the suggestion that
initial neutron star spin periods depend on their
magnetic fields; in particular, there is a tendency for high-field
systems to be born as slow rotators.  The aim of this
project is to understand the emission properties of HBRPs,
using multiple radio frequencies and high time resolution data. 
One specific objective is to
identify HBRPs radio emission characteristics that are different
from those of normal pulsars.
\end{abstract}

\section{Introduction}

\begin{table*}[t]
\caption{Technical details for both modes of observations (multifrequency 
polarisation and single pulse). We used two receivers at Parkes radio 
telescope: MULTI -- centre beam  at the 13-multi-beam receiver and 1050CM -- 
10 cm and 50 cm bands at 10/50cm receiver, with different backend 
configurations: CPSR2 -- Caltech-Parkes-Swinburne baseband Recorder II; 
PDFB1 -- Prototype Pulsar Digital Filter Bank, WBCORR -- Wideband Correlator, 
FB -- Filterbank.}
\label{table1}
\centering
\begin{tabular}{l l l l l | l } 
\hline \hline
  & \multicolumn{4}{l}{Polarisation at } & \multicolumn{1}{r}{Single pulse} 
\\ \hline
Center Frequency [MHz] &  685   & 1369  & 1433 & 3100 & 1374   \\
Instrument name: & CPSR2 & PDFB1 & WBCORR & PDFB1 & FB \\
Nr of bins in profile & 1024 & 512 & 2048 & 512 & 256 \\
Bandwidth [MHz] & 64 & -256 & -256 & 256 & -288 \\
Receiver name & 1050CM & MULTI & MULTI & 1050CM & MULTI \\ \hline \hline
\end{tabular}
\end{table*}
 
To obtain a good understanding of the physics of the pulsar
magnetosphere, we need information on time-averaged polarisation
profiles, which tell us about the structure of the magnetic field, the
properties of the magnetosphere and the geometry of the star, and on
single pulse profiles, which give us information on the instantaneous plasma
conditions and radiation mechanism. To understand emission
constraints from the HBRPs\footnote{The B-field in some radio pulsars is 
stronger  than the electron critical field ($\sim 4\times 10^{13}\rm\,G$), 
at which the cyclotron energy of an electron rotating around the
magnetic field line reaches its rest mass energy.} we observed 34 long-period 
pulsars
(17 HBRPs and 17 low-magnetic-field pulsars) at three different
frequencies (in the range 700 - 3100 MHz) in order to achieve
valuable multifrequency polarisation profiles, spectral indices values
and single pulse profiles. Table~\ref{table1} summarises observational 
characteristics for both modes of observations (multifrequency polarisation 
and single pulse, respectively) 
of the technical details for two receivers and different configurations used 
during our observations. The first and second rows in the table list central 
frequencies and  backend instrumentation used; the next row lists the 
observational resolution, which is  given by the number of bins in the 
profile; the observational bandwidths and used receivers are listed in the 
fourth and the fifth rows. \\
Our aims in this project are: 1) to investigate if HBRPs form transition 
objects between the normal
pulsar population and magnetars or if they form a separate pulsar
population, and 2) to understand recent results from our paper 
Vranesevic et al. (2004):\nocite{vml+04a} a) why HBRPs contribute half 
to the 
total pulsar
birthrate (even though they contribute to only few per cent of the
total pulsar population), and b) why up to 40\,\% of all pulsars are born
with periods in the range 100--500\,ms (which is contrary to
the usual view that all pulsars are born as fast rotators). Here
we present first results on the multi\-frequency polarimetry for the
most interesting pulsars from our samples.

\section{The first results}
  
For the purpose of the Bad Honnef Meeting we present the first results for 
five  pulsars: four HBRPs (PSR J1718-3718, PSR J1734-3333, PSR J1814-1744, \& 
PSR J1847-0130) plus a representative of our low-magnetic field pulsar sample, 
PSR J2144-3933. In Table~\ref{table2} pulsar parameters are listed: pulsar 
spin characteristics, their Galactic positions, dispersion measures, widths of
 pulse at 50\% of peak, mean flux densities at 1400 MHz, pulsar distances, 
spin down ages, surface magnetic flux densities and spin down energy loss 
rates.
The detection of these pulsars was challenging because: a) detection of 
long-period pulsars using
conventional search techniques is hard due to red noise in the Fourier 
transform of
time series seriously reducing the sensitivity; b) all our pulsars are
very faint, near the detection limit of the instrument; and c) for the 
generally accepted spectral index of -1.8 we would need hundreds of
hours of observations to achieve a significant signal to noise at 3100 MHz. 
 
The fact that we detected HBRPs at 3100 MHz suggests that they must have 
relatively flat radio spectra. Data at 700 MHz are still being analysed. 

\begin{figure}
\begin{center}
\begin{tabular}{l l}
 a) & b) \\
\includegraphics[width=40mm,height=40mm,angle=270]{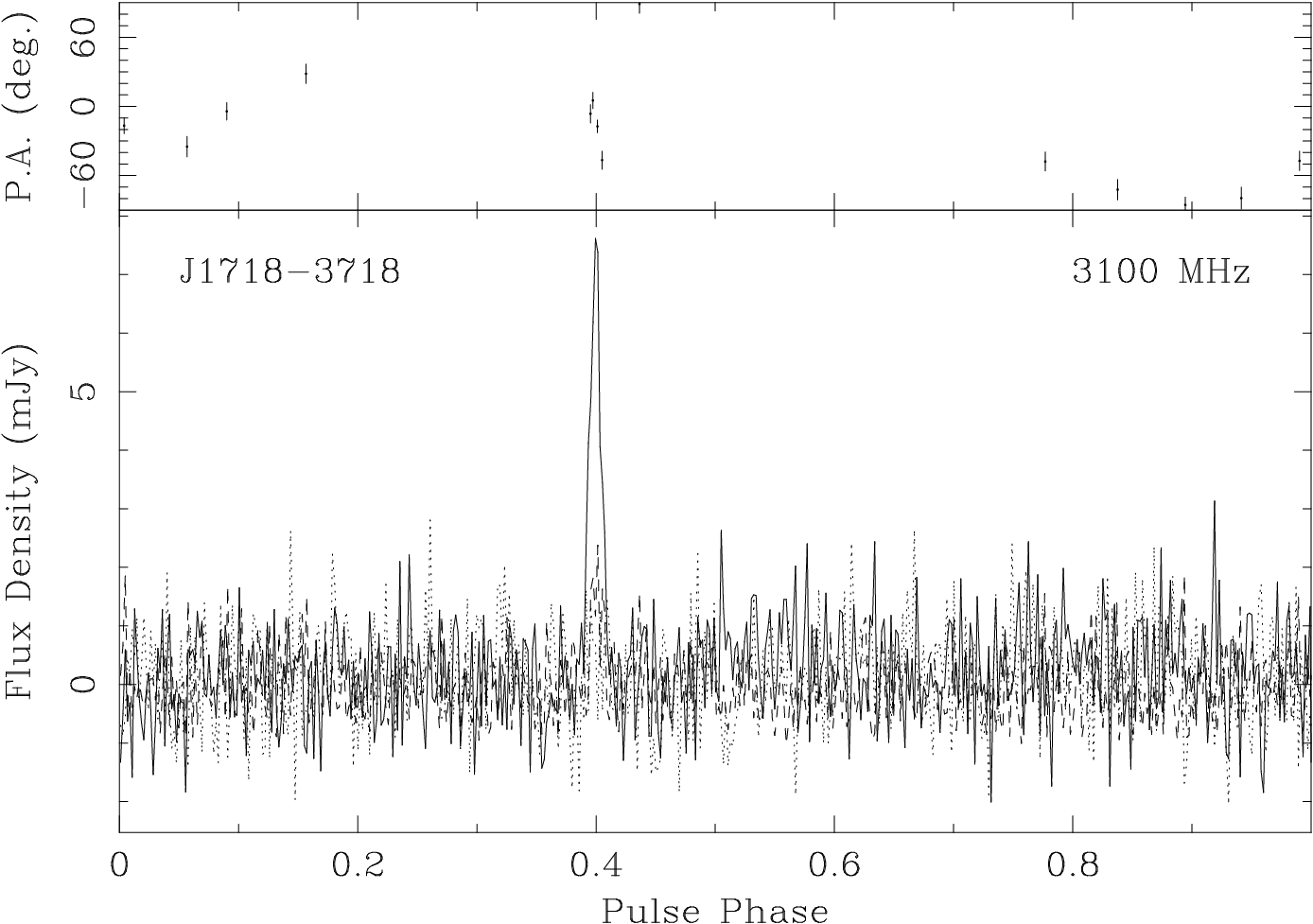} & 
\includegraphics[width=40mm,height=40mm,angle=270]{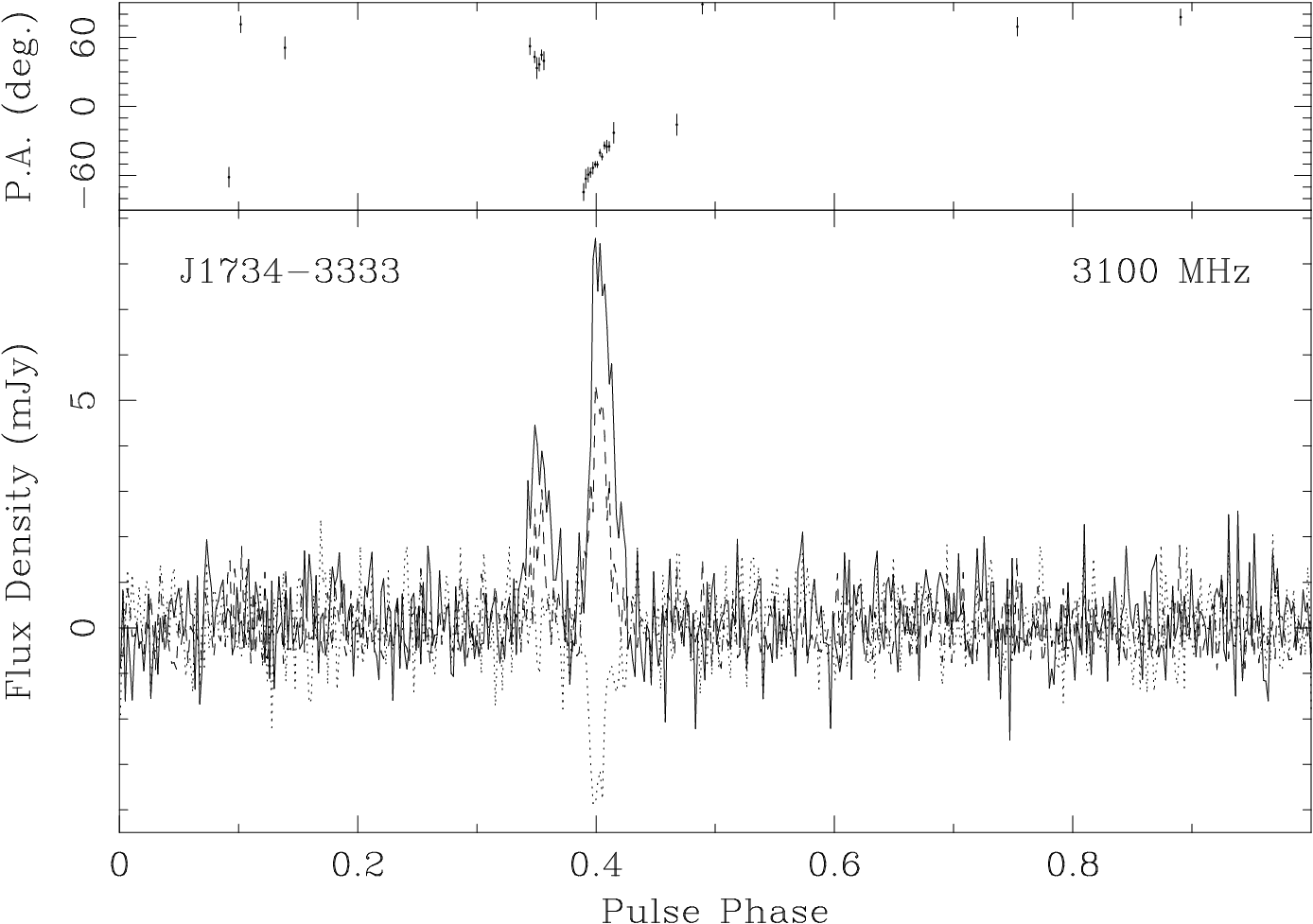} \\
\includegraphics[width=40mm,height=40mm,angle=270]{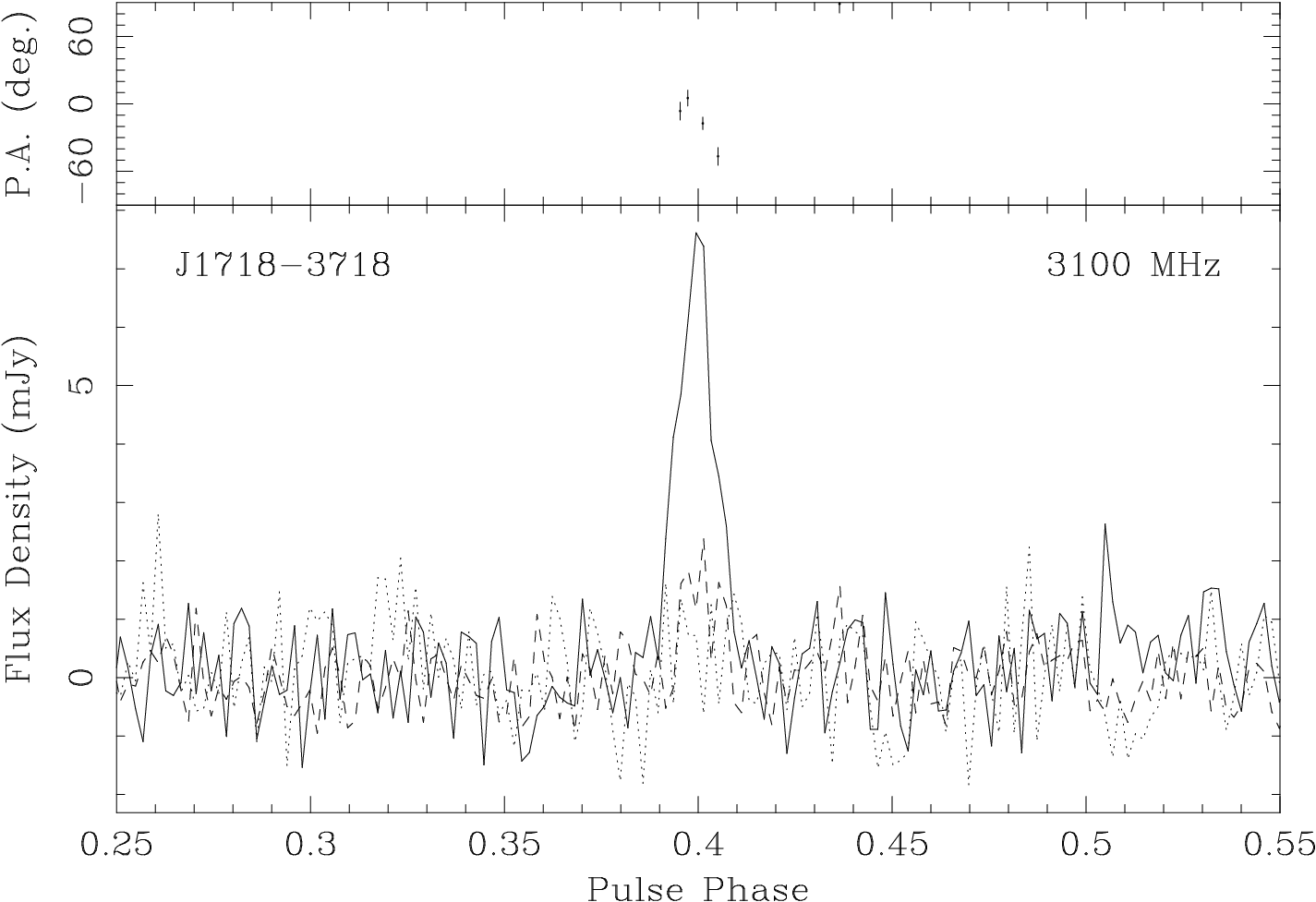} &
\includegraphics[width=40mm,height=40mm,angle=270]{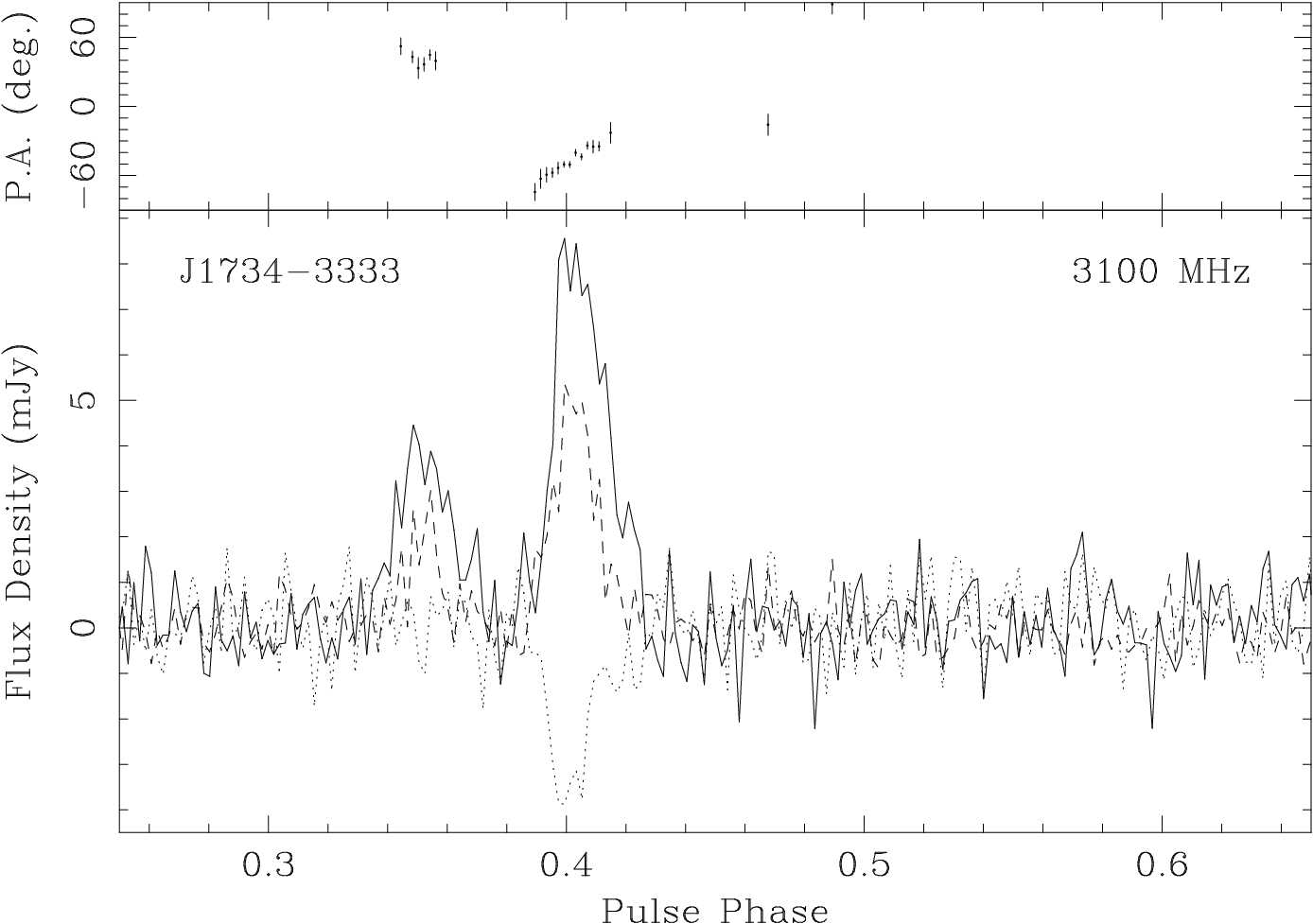} \\
\includegraphics[width=40mm,height=40mm,angle=270]{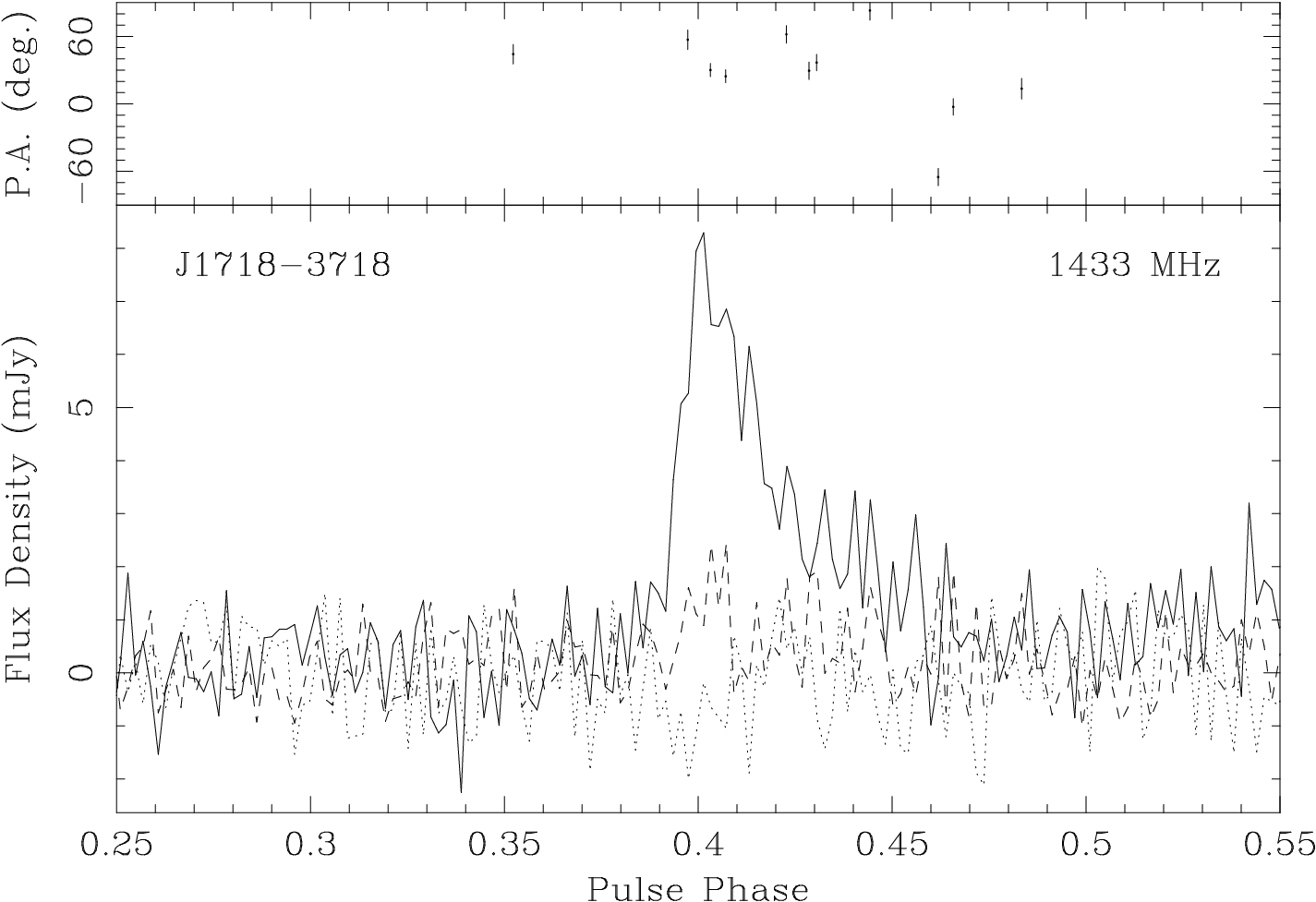} &
\includegraphics[width=40mm,height=40mm,angle=270]{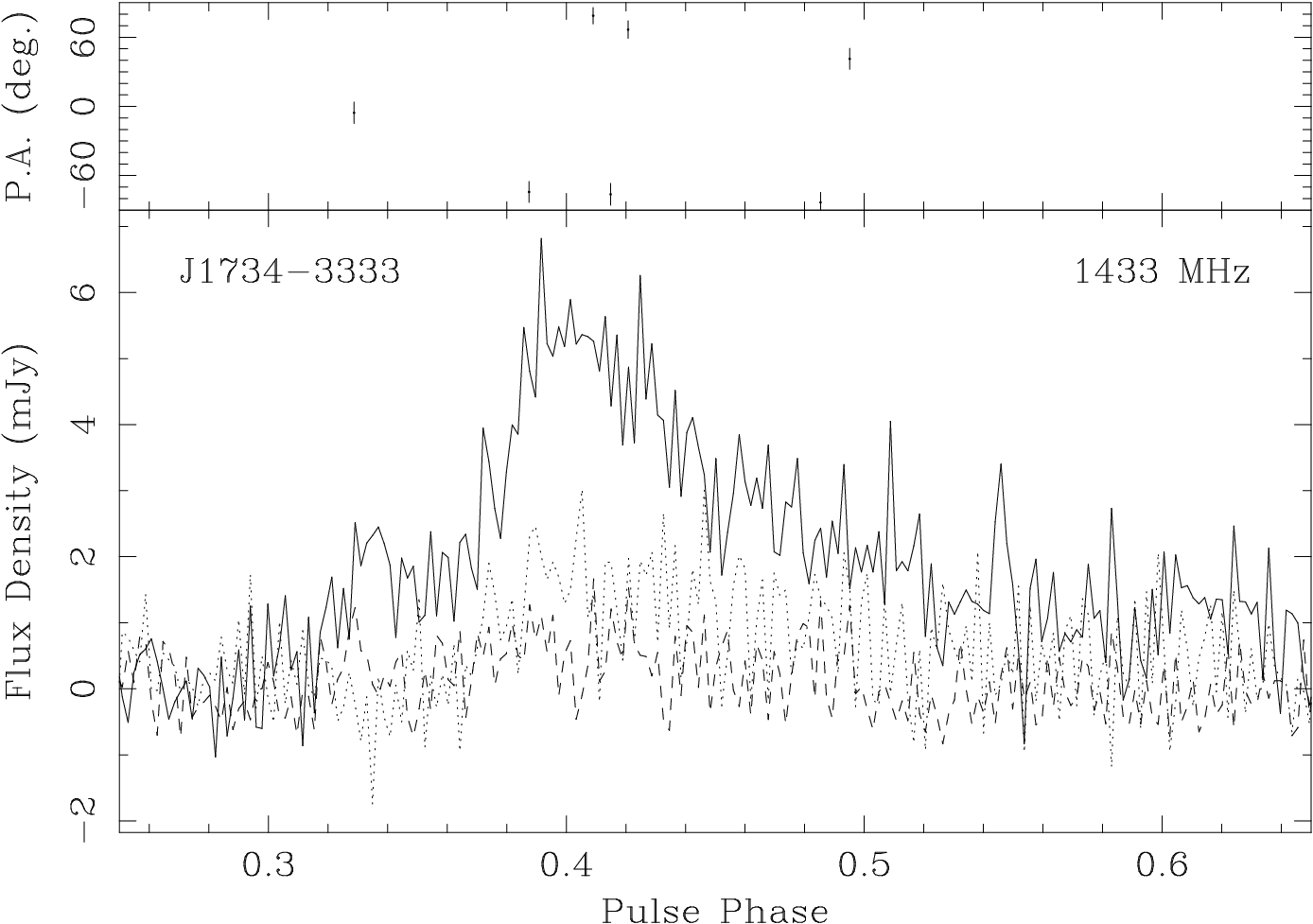} \\
 \end{tabular}
\caption{Multifrequency polarisation profiles for pulsars: a) PSR J1718-3718 
and b) PSR J1734-3333. The top panels show integrated pulse profiles plotted 
for a whole pulsar period versus flux at 3100 MHz. The middle and bottom 
panels plot  zoom in pulsar profiles at 3100 MHz \& 1433 MHz. All panels show 
total intensity as a solid line, with linear intensity as dashed, and circular
 intensity as dotted lines. The upper frames for all panels show the position 
angle  of the linear polarisation.
\label{image1}}
\end{center}
\end{figure}

\begin{figure}
\begin{center}
\begin{tabular}{l l}
 a) & b) \\
\includegraphics[width=40mm,height=40mm,angle=270]{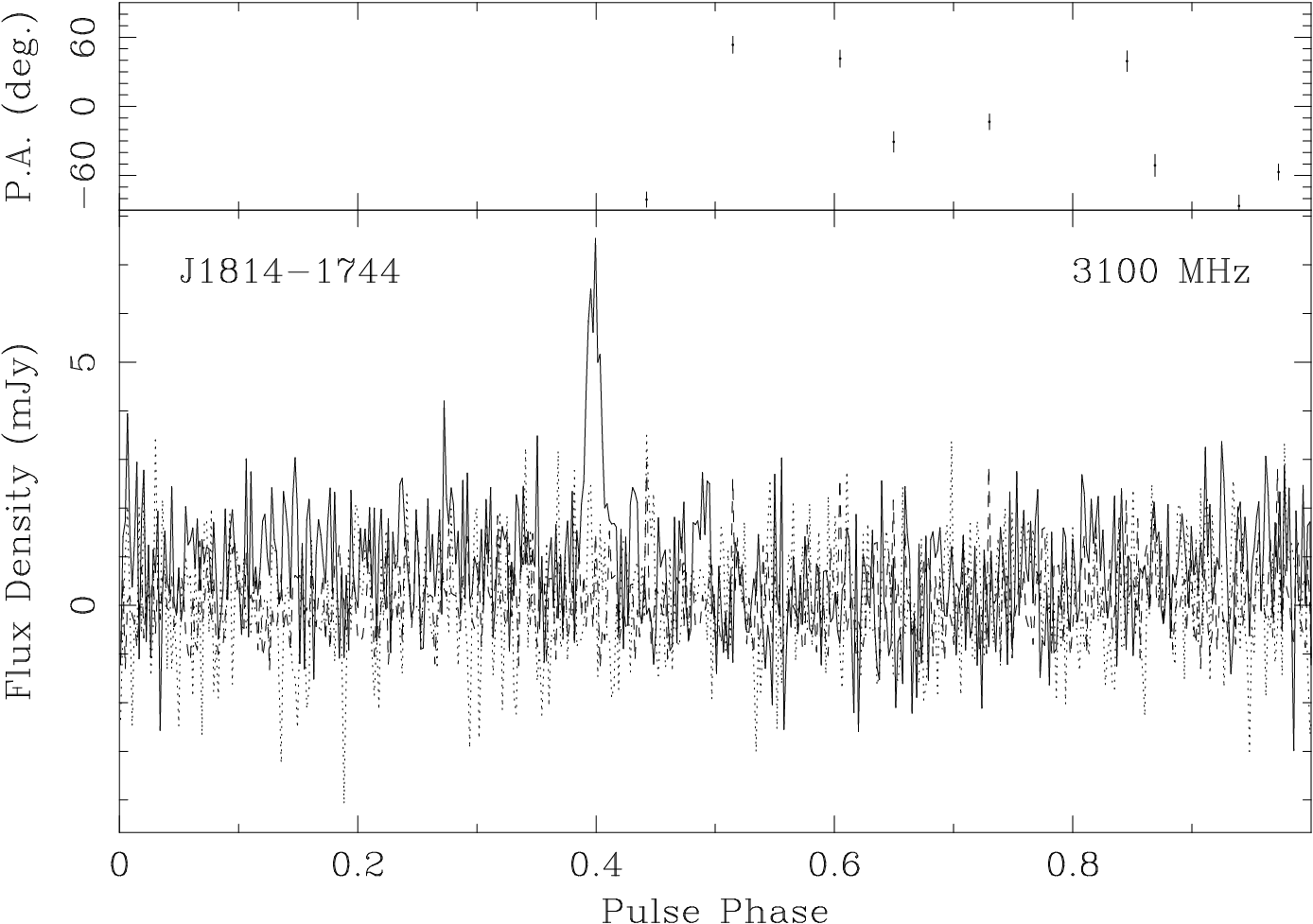} &
\includegraphics[width=40mm,height=40mm,angle=270]{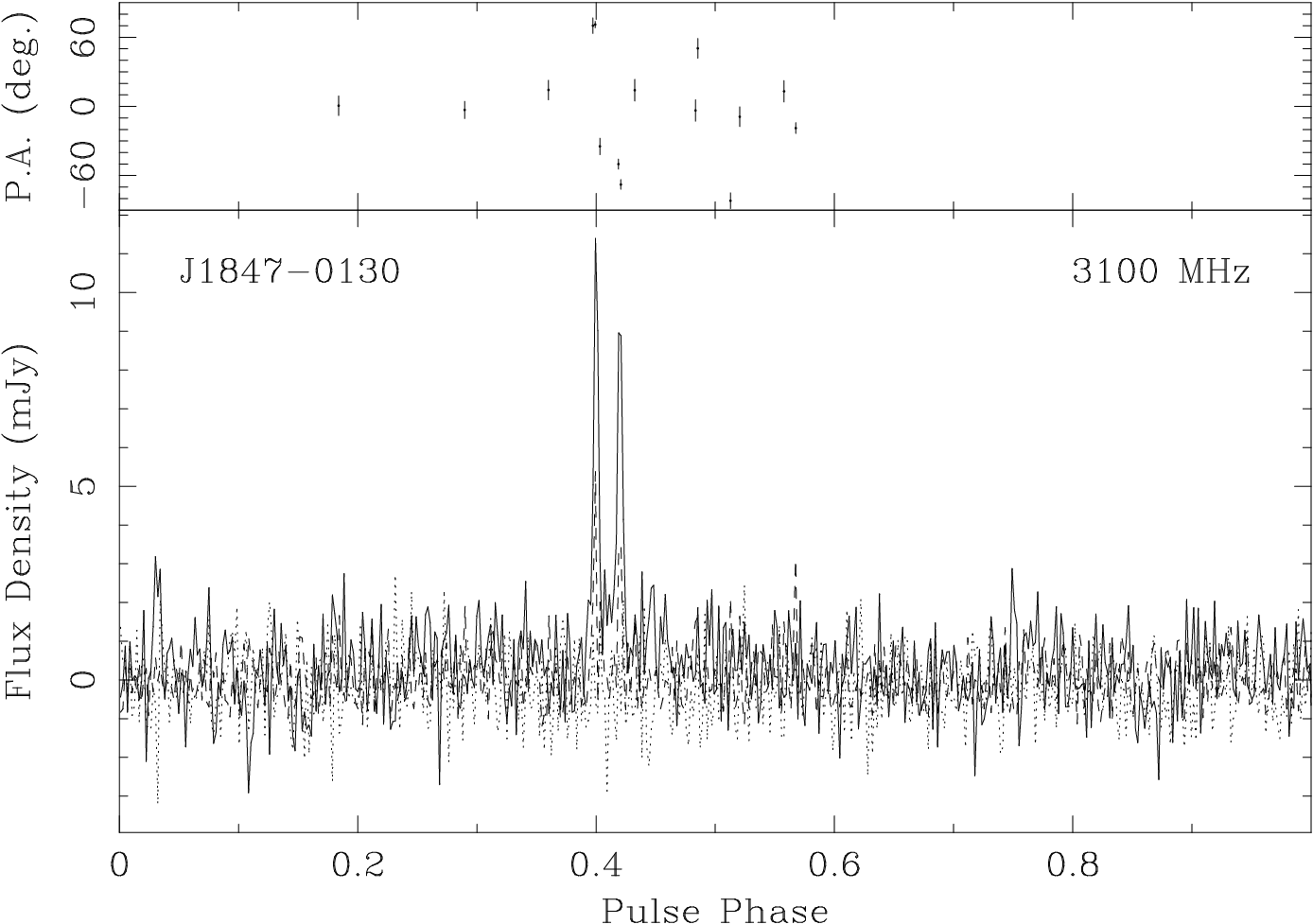} \\
\includegraphics[width=40mm,height=40mm,angle=270]{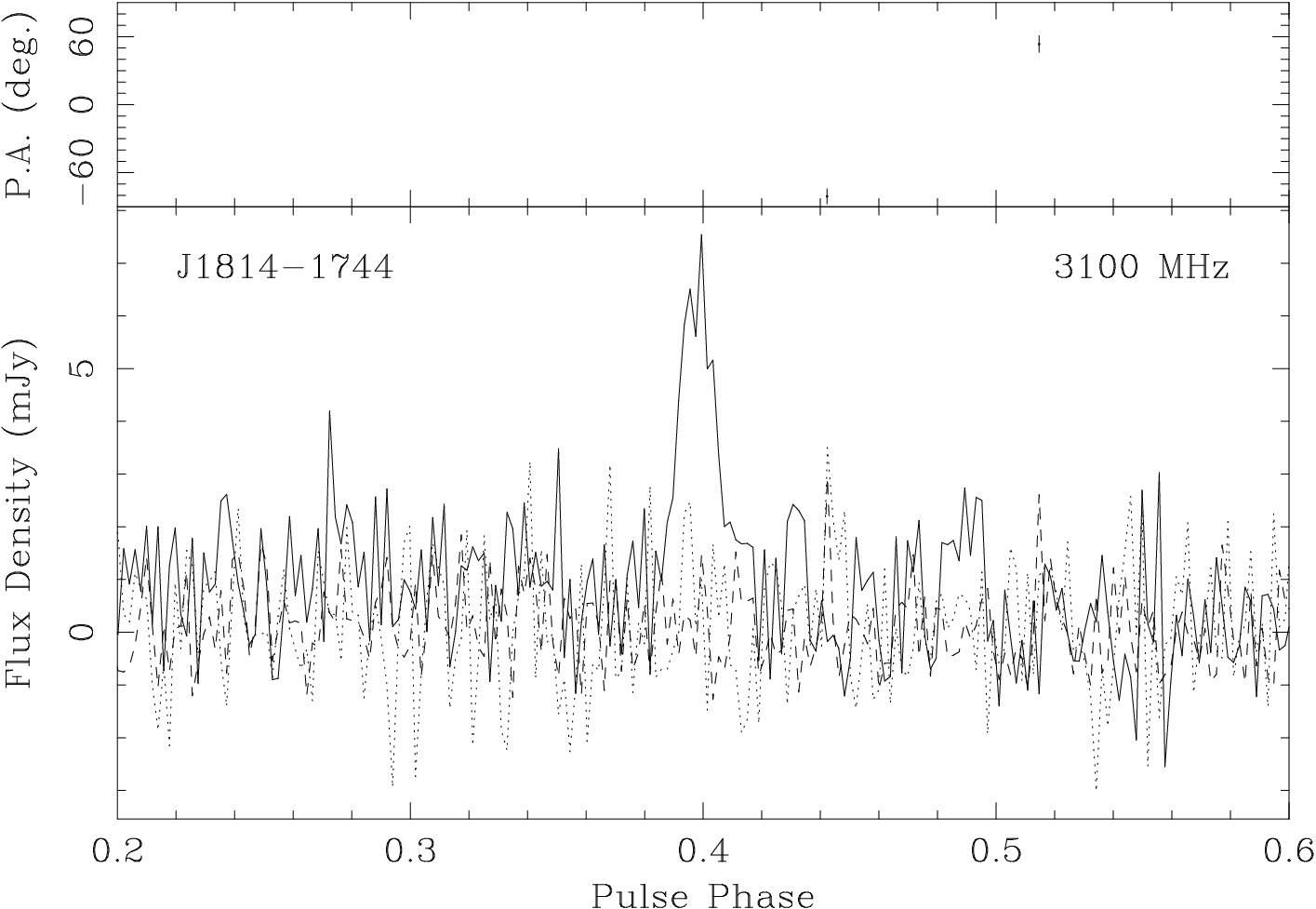} &
\includegraphics[width=40mm,height=40mm,angle=270]{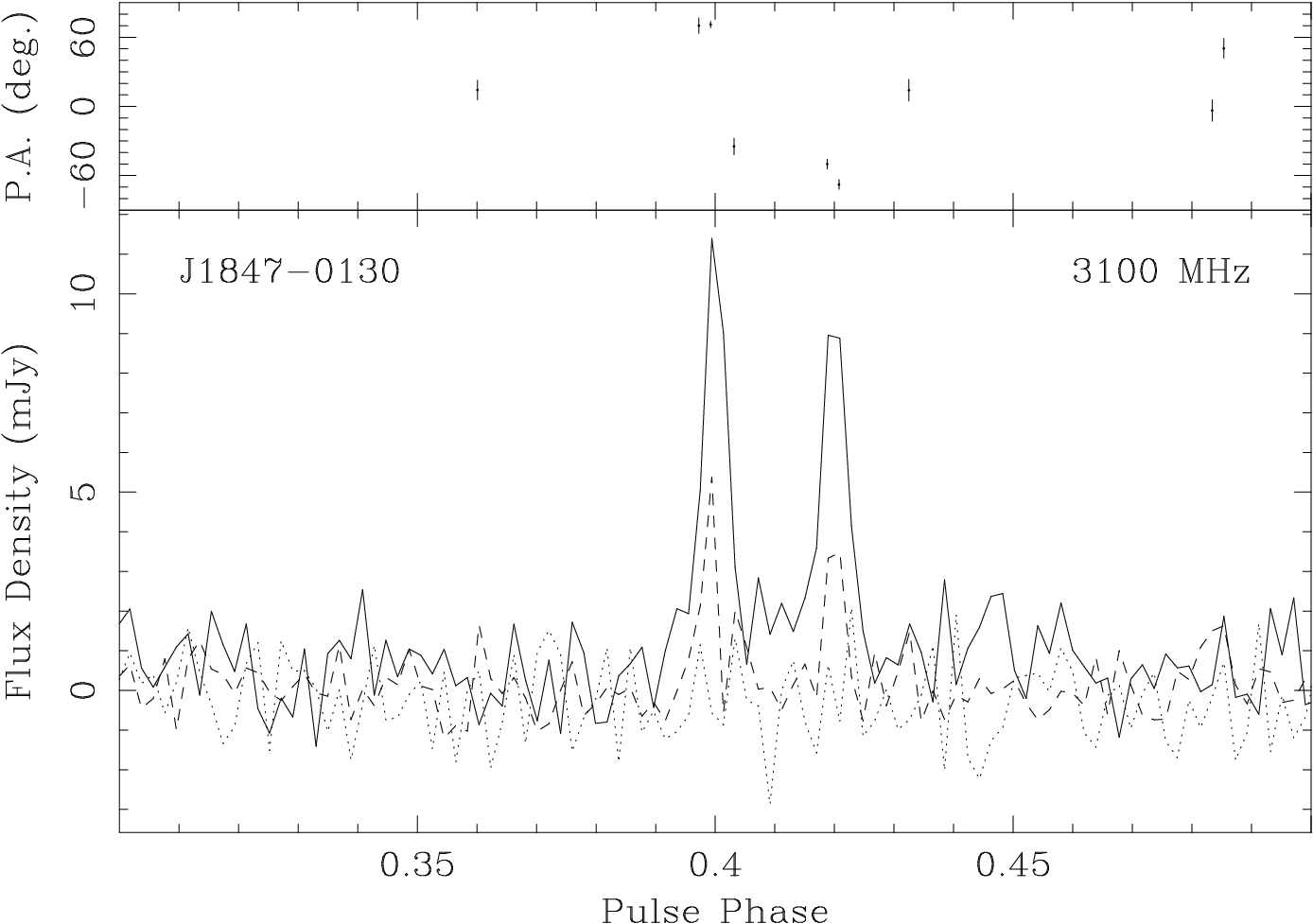} \\
\includegraphics[width=40mm,height=40mm,angle=270]{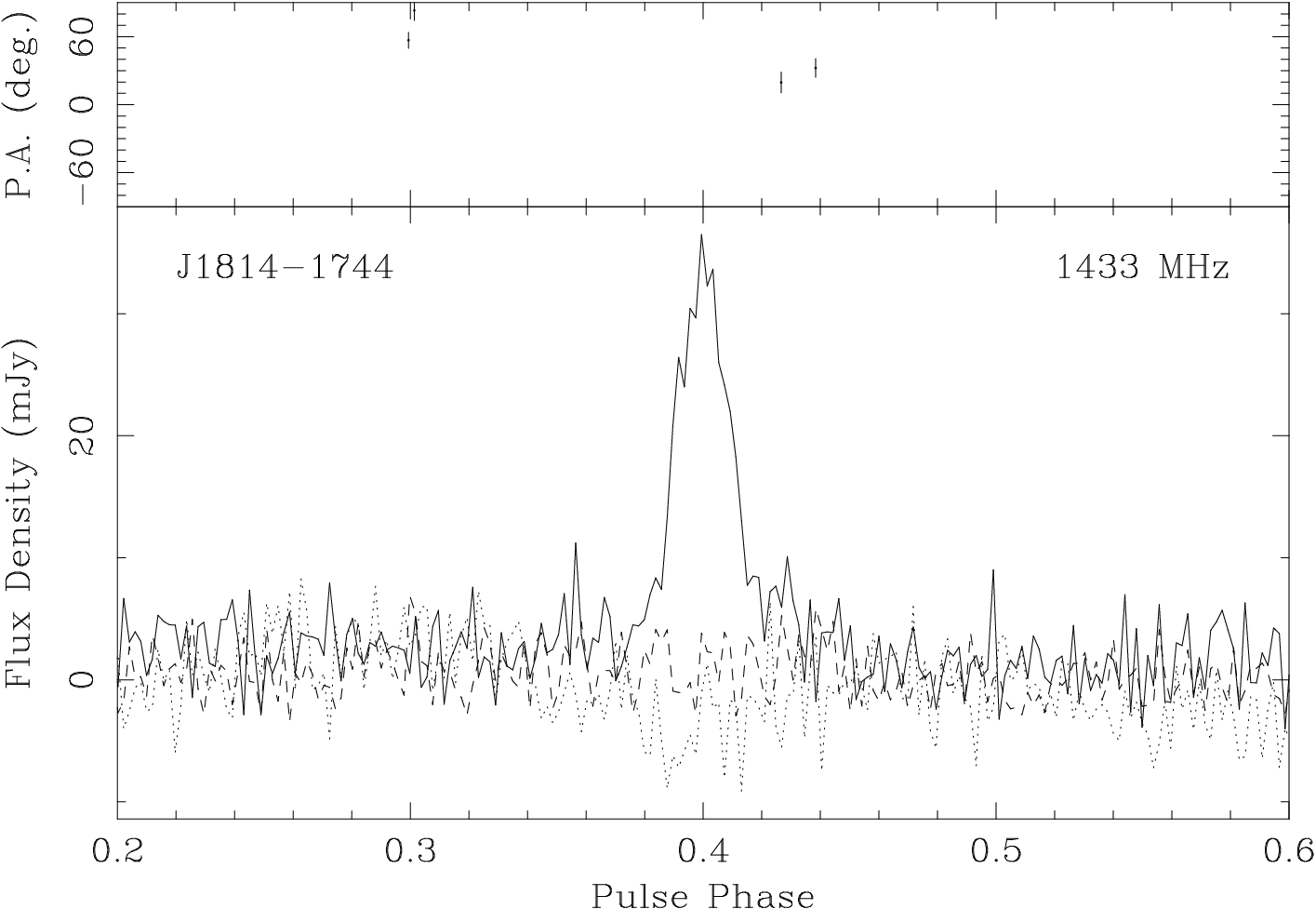} &
\includegraphics[width=40mm,height=40mm,angle=270]{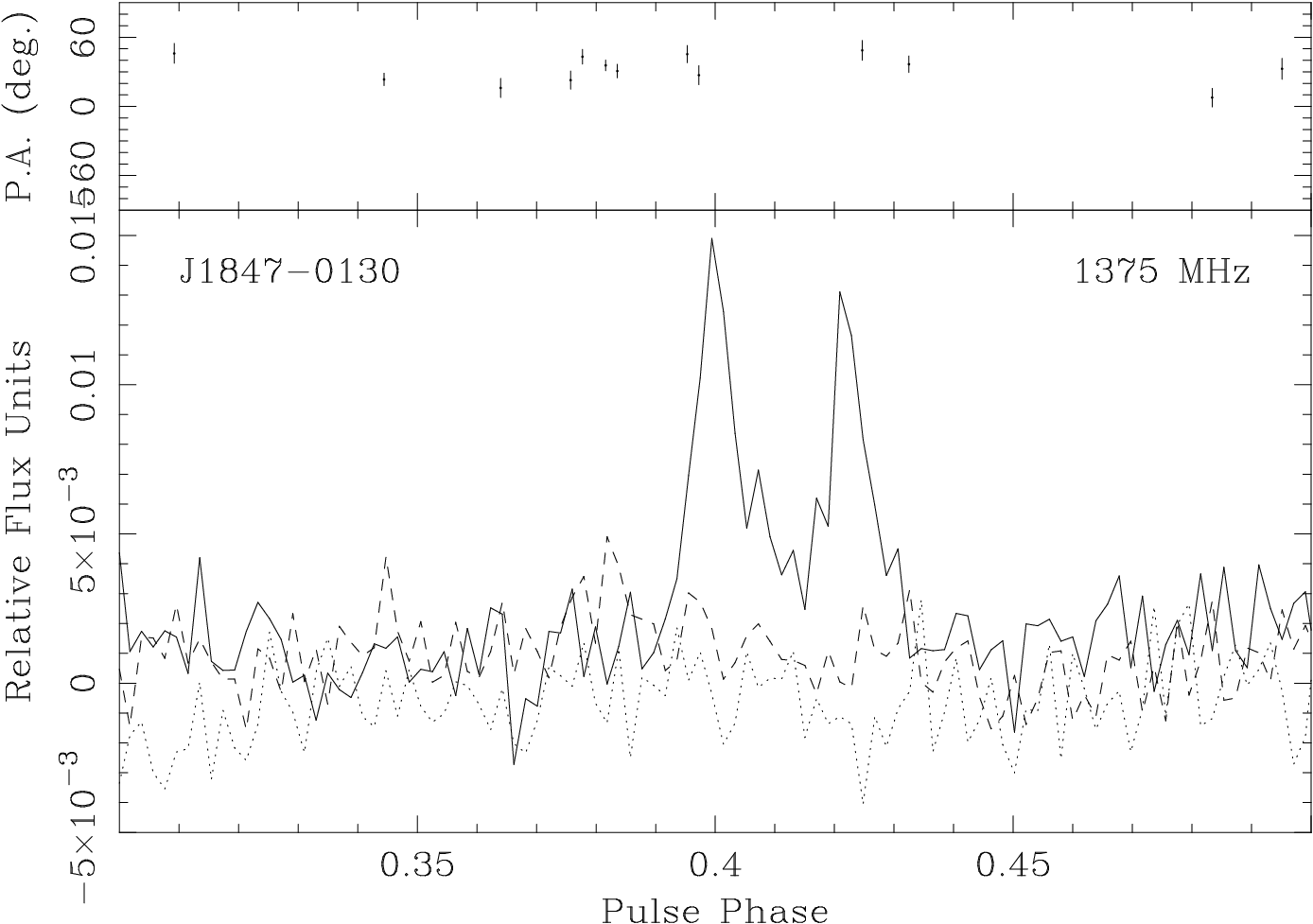}\\
\end{tabular}
\caption{Multifrequency polarisation profiles for pulsars: a) PSR J1814-1744 
and b) PSR J1847-0130. The top panels show integrated pulse profiles plotted 
for a whole pulsar period versus flux at 3100 MHz. The middle and bottom 
panels plot  zoom in pulsar profiles at 3100 MHz \& 1433 MHz for a) and 1375 
MHz for b). All panels show total intensity as a solid line, with linear 
intensity as dashed, and circular intensity as dotted lines. The upper frames 
for all panels show the position angle  of the linear polarisation.
\label{image2}}
\end{center}
\end{figure}

\begin{figure}
\begin{center}
\begin{tabular}{l l}
a) & b) \\
 \includegraphics[width=40mm,height=40mm,angle=270]{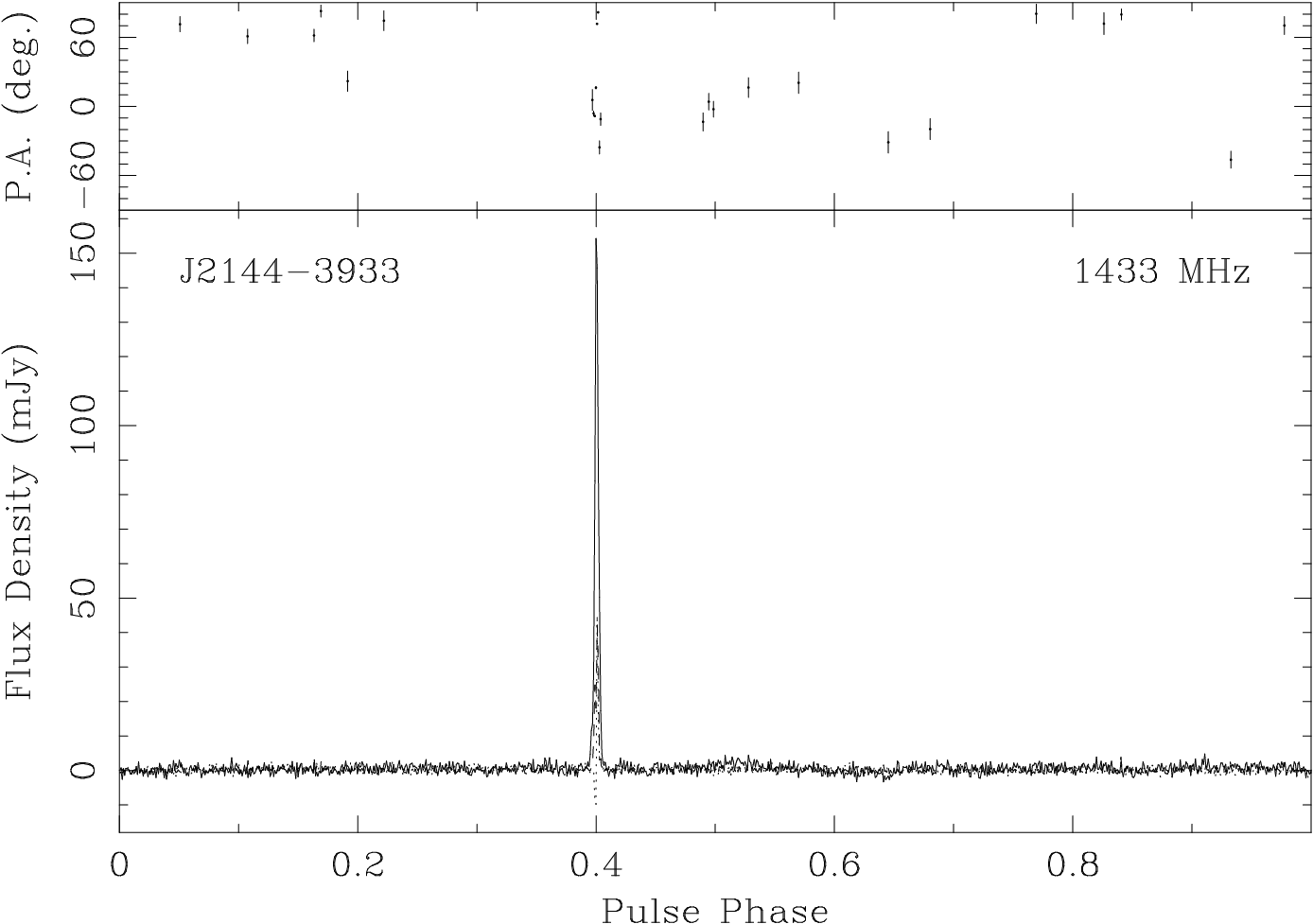} &
  \includegraphics[width=40mm,height=40mm,angle=270]{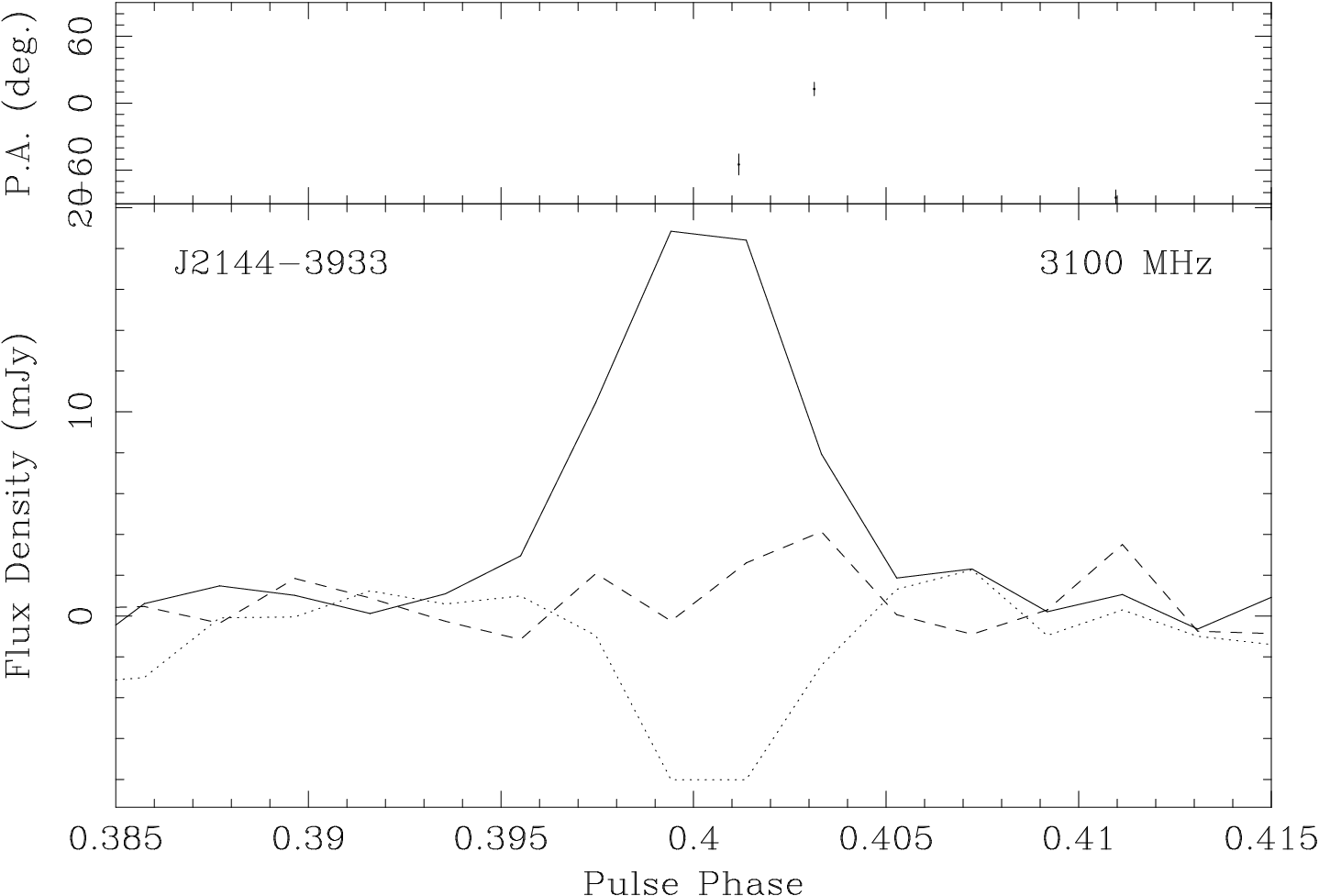} \\
  c) & d) \\
 \includegraphics[width=40mm,height=40mm,angle=270]{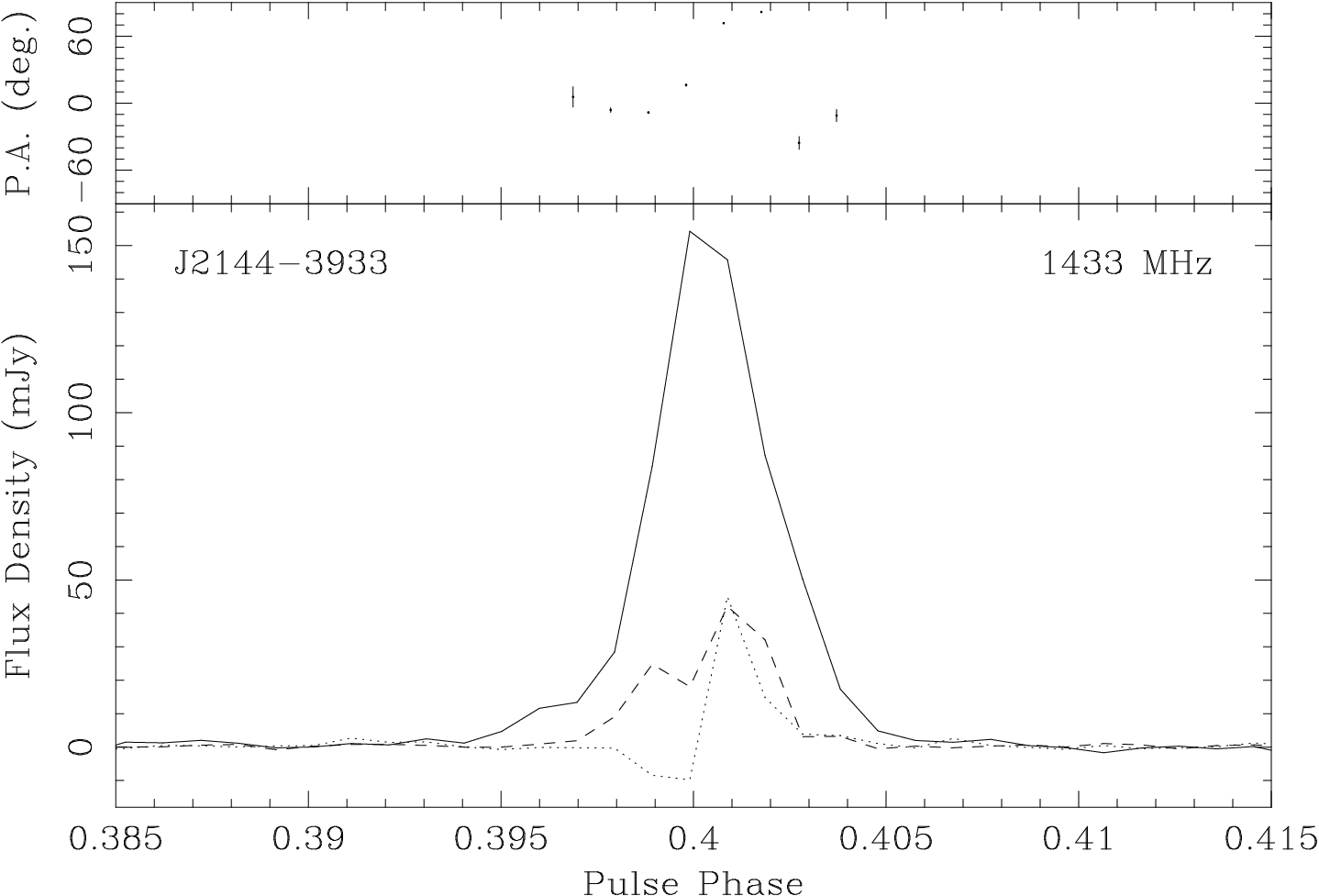} & 
 \includegraphics[width=40mm,height=40mm,angle=270]{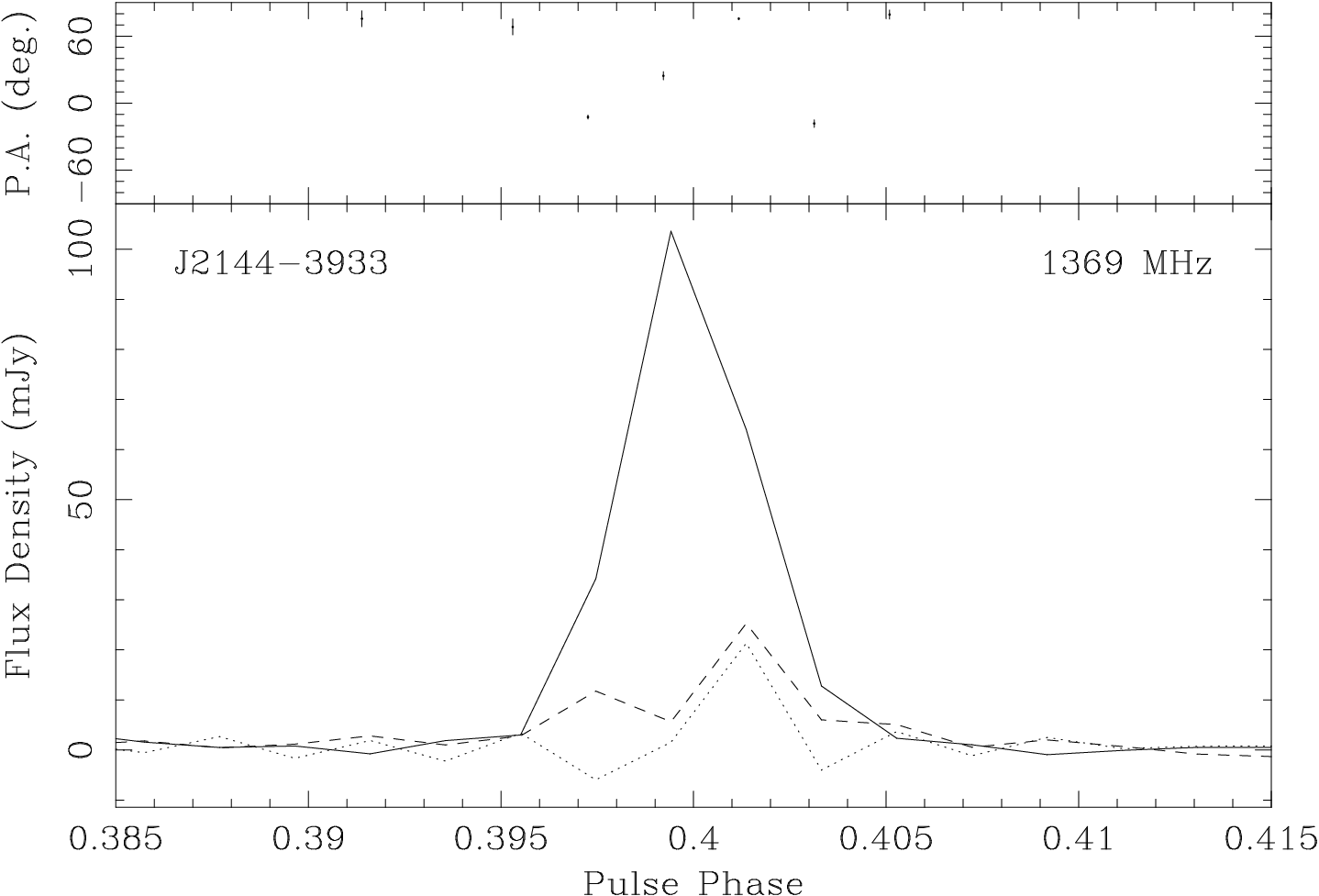} \\
 \end{tabular}
\caption{Multifrequency polarisation profiles for pulsar PSR J2144-3933, 
showing total intensity as a solid line, with linear intensity as dashed, and 
circular intensity as dotted lines; the upper frames for all panels show the 
position angle  of the linear polarisation. The panel a) shows integrated 
pulse profiles plotted for a whole pulsar period versus flux at 1433 MHz. 
Panels b), c), and d) plot  zoom in pulsar profiles at 3100, 1433 and 1369 
MHz, respectively.
  \label{image3}}
\end{center}
\end{figure}

Profiles at 1433 MHz for pulsars PSR J1718-3718 and PSR J1734-3333 
show significant broadening of an intrinsically
sharp pulse (compare middle and bottom panels of Figure~\ref{image1}), which 
is due to ray scattering by irregularities in the
ISM (characteristic of distant, high DM pulsars). The other two HBRPs shown 
here, PSR J1814-1744 and PSR J1847-0130, are also very distant with high DMs 
(Table~\ref{table2}), but do not show significant
scattering. This is consistent with greater scattering along lines of
sight that pass nearer to the Galactic centre.
For most pulsars the
fractional linear polarisation decreases with increasing
frequency. However, the HBRPs shown here
(except J1814-1744, which has no detectable polarisation,
see Figure~\ref{image2}a) have higher linear polarisation at 3100 MHz than at
lower frequencies. All these HBRPs are young pulsars with pulse
profiles which consist of one or two prominent components with higher
linear polarisation at higher frequency, which is consistent
with the recent results by Johnston \& Weisberg (2006).\nocite{jw06} \\
PSR J2144-3933 which is part of our
low-magnetic-field sample, shows curious features, see  Figure~\ref{image3}: 
a) polarisation
intensities at low frequencies are stronger compared with published
data \cite{mhq98}, b) there is a significant 
increase in circular polarisation
going towards higher frequencies, and c) depolarization of linear
component at 3100 MHz (which may be due to reduction of profile
resolution). All of these are going to be explored in more detail.

\begin{table*}[t]
\caption{Pulsar profile and timing solution parameters, position and some 
derived parameters for four HBRPs and one low-magnetic field pulsar.}
\label{table2}
\centering
\begin{tabular}{c c c c c c c c c c c c } 
\hline \hline
 & $P$ & $\dot{P}$  & l & b & DM & $W_{50}$ & $S_{1400}$ & d & $\tau_c$  & 
$B_{surf}$ & $\dot{E}$ \\
 & $[\rm\,s$] & $[10^{-15}\rm\,s\,s^{-1}]$ &  [$^{\circ}$]  &  [$^{\circ}$] & 
[$\rm\,cm^{-3}\,pc$]  &  [$\rm\,ms$]  &  [$\rm\,mJy$]  & [$\rm\,kpc $]  &  
[$\rm\,kyr$] &   [$10^{13}\rm\,G$] &  [$10^{32}\rm\,ergs\,s^{-1}$] \\ \hline 
 PSR: & \\
 \multicolumn{12}{c}{HBRPs} \\ 
J1718-3718 &  3.4  & 1600 & 350  & 0.22 & 373 & 130 & 0.18  & 4.86 & 33.5  & 
7.44  & 16 \\
J1734-3333  &  1.169 & 2280 &  354.82 &  -0.43 &  578 &  164.1 &  0.5 &  7.40 
& 8.13 & 5.22 & 560 \\
J1814-1744  &  3.98 &  743 &  13.02 &  -0.21 &  792 &  92.0 &  0.7 &  9.76 &  
84.8 &  5.5 &   4.7 \\
J1847-0130 & 6.707 &  1270 &   31.15 &  0.17 &  667 &  205 &  0.33 &  7.74 &  
83.3 &  9.36 & 1.7 \\ \\
 \multicolumn{12}{c}{Low-field pulsar}  \\ \
J2144-3933 & 8.5 &  0.496 &  2.79 & -49.47 & 3.35 & 25 & 0.8 & 0.18 & 272000 & 
0.208 & 0.00032 \\ \hline \hline
\end{tabular}
\end{table*}

\section{Discussion}
 
Using the birthrate code from Vranesevic et al. (2004) and accurately
accounting for all known selection effects and using the beaming
fraction given by Tauris \& Manchester (1998)\nocite{tm98}, 
we calculated that 187$\pm$103 long-period radio
pulsars with magnetic field above the quantum critical field are
active in the Galaxy and that one such pulsar is born each 500
years. 

It is puzzling that this calculated number of HBRPs in the
Galaxy is comparable with the predicted number of neutron stars at the
supersonic propeller stage, according to the results presented by
Beskin at the meeting. In their analysis of statistical distribution of
extinct radio pulsars \cite{be05a} and neutron stars at the supersonic
propeller stage \cite{be05b}, they include evolution of the axial
inclination and use two models for the particle acceleration region:
hindered particle escape from the stellar surface \cite{rs75}, and
free particle escape \cite{aro79}.  They found that transition of a
radio pulsar to the propeller stage can occur at the short periods
$P\sim 5-10\rm\,s$ and the number of those extinct radio pulsars (with
spin parameters similar to HBRPs) is much larger than when using
standard model (in which no evolution of inclination angle of magnetic
axis to the spin has been accounted for).

Another interesting result regarding the influence of inclination
angle $\alpha$ is on the stability of dipolar magnetostatic
equilibrium in newly born neutron stars, presented at the meeting by
Geppert \& Rheinhardt. They assumed that newly born NSs with highly
magnetized progenitors and proto-NS phase surface magnetic fields of
$\ga10^{15}\rm\,G$ (gained by flux conservation) that also have
sufficiently fast rotation (initial period of $P\la6 $ms) and $\alpha
\la 45^{\circ}$ retain their magnetic fields and appear, after a rapid
spin down, as magnetars. Others (with $P\ga6\rm\,ms$ and $\alpha \ga
45^{\circ}$) lose almost all of their initial magnetic energy by
transferring it into magnetic and kinetic energy of relatively
small-scale fields and continue their life as radio pulsars with a
dipolar surface field of $10^{12-13}$G (for more details see Geppert
\& Rheinhardt 2006\nocite{gr06}). The implication for HBRPs, as a
young highly magnetised objects, is that they should have
$\alpha<45^{\circ}$. This is the case for one of the well known HBRPs,
the 1610-year-old pulsar PSR J1119-6127, $\alpha=19^{\circ}$
\cite{gr06}. Gonzalez, at this meeting, presented recent results on
X-ray detection from this high-B radio pulsar and from a few other
HBRPs, with the main highlights being their unusual thermal emission,
which they explained in terms of anisotropic high temperature
distribution and a small emitting area.

The detection of magnetospheric radio emission from a magnetar has just been 
announced by Camilo et al. (2006)\nocite{crh+06}. The data show highly 
linearly polarized, bright 
radio pulsations from XTE J1810-197, which is a transient AXP. The fact that 
this source had very faint X-ray properties in its quiescent phase, similar 
to soft X-ray detection of PSR J1718-3718, suggests that XTE J1810-197 could 
have generated familiar radio pulsar emission before the outburst observed in 
early 2003. This makes a plausible direct link between radio pulsars and 
magnetars.       

\section{Summary}

The idea that HBRPs were
born as a slow rotators is in agreement with Ferrario \& Wickramasinghe (2006)
\nocite{fw06}, who
argued that initial neutron star spin periods depend
critically on their magnetic fields. There is a
tendency for high-field systems to be born as slow rotators. 
Recent results by Vink \& Kuiper (2006)\nocite{vk06} show no evidence for 
magnetars being
formed from millisecond proto-NSs. It is possible that
magnetars have stellar progenitors with high magnetic field
cores, which is the fossil-field hypothesis of Ferrario \& Wickramasinghe 
(2006) \nocite{fw06}. It
remains unclear whether HBRPs evolve into magnetars or whether HBRPs and
magnetars form distinct populations from birth. 

\begin{acknowledgements}
NV gratefully acknowledges the support by the ATNF Graduate Student Research 
Scholarships for the Bad Honnef meeting trip. The ATNF Pulsar Catalogue has 
been used: {\it http://www.atnf.csiro.au/research/pulsar/psrcat/}, 
\cite{mhth05}. All data were analysed  off-line using the {\it PSRCHIVE } 
software package, \cite{hvm04}: {\it http://astronomy.swin.edu.au/pulsar/}. 
The Parkes telescope is part of the Australia Telescope which is funded by 
the Commonwealth of Australia for operation as a National Facility managed by 
CSIRO. 
\end{acknowledgements}
   

\bibliographystyle{aa}
\bibliography{journals,modrefs,psrrefs,crossrefs,nvrefs}

    \clearpage

\end{document}